\documentclass[twocolumn,showpacs,preprintnumbers,amsmath,amssymb,superscriptaddress]{revtex4}

\usepackage{graphicx}
\usepackage{dcolumn}
\usepackage{bm}

\begin{document}

\title{Bistability of optically-induced nuclear spin orientation in quantum dots}

\author{A. Russell}
\affiliation{Department of Physics, University of Lancaster,
Lancaster, LA1 4YB, UK}
\author{Vladimir I. Fal'ko}
\affiliation{Department of Physics, University of Lancaster,
Lancaster, LA1 4YB, UK}
\author{A. I. Tartakovskii}
\affiliation{Department of Physics and Astronomy, University of
Sheffield, S3 7RH, UK}
\author{M. S. Skolnick}
\affiliation{Department of Physics and Astronomy, University of
Sheffield, S3 7RH, UK}

\date{\today}

\begin{abstract}

We demonstrate that bistability of the nuclear spin polarization in
optically pumped semiconductor quantum dots is a general phenomenon
possible in dots with a wide range of parameters. In experiment,
this bistability manifests itself via the hysteresis behavior of the
electron Zeeman splitting as a function of either pump power or
external magnetic field. In addition, our theory predicts that the
nuclear polarization can strongly influence the charge dynamics in
the dot leading to bistability in the average dot charge.
\end{abstract}

\pacs{73.21.La, 72.25Fe}

\maketitle

The hyperfine interaction in solids between the electron and nuclear
spins \cite{overhauser} leads to the Overhauser energy shift
$\delta$ of the electron spin states, produced by the magnetic
dipole moments of orientated nuclear spins, often described in terms
of the effective nuclear magnetic field
$B_{N}=2\delta/g_{e}\mu_{B}$. The hyperfine interaction is also
responsible for the transfer of spin from electrons to nuclei and
may lead to a significant nuclear spin polarization $S$, if the
system is pumped with highly polarized electrons
\cite{pines,paget,braun}. Recently, nuclear spin effects have been
observed in optically pumped semiconductor quantum dots (QDs)
\cite{gammon,yokoi,lai,akimov,brauntwo,tartakovskii,maletinsky,eble}.
In these experiments circularly polarized light generated
electron-hole pairs which, then, relax into the ground state of the
dot, with electrons exhibiting a longer spin memory than the holes
(which lose their spin polarization due to a stronger spin-orbit
coupling \cite{uenoyama,ebbens}).

Recently, the nuclear spin orientation in optically pumped dots has
been found to display a pronounced bistability in externally applied
magnetic fields \cite{brauntwo,tartakovskii,maletinsky}. This
appears as a threshold-like switching of the nuclear magnetic field
$B_N \sim 2-3T$ and a characteristic hysteresis behavior observed in
the dependence of the nuclear polarization on either the intensity
of the polarized light \cite{brauntwo,tartakovskii} or external
magnetic field \cite{maletinsky,ono}.

\begin{figure}
\includegraphics[width=8cm]{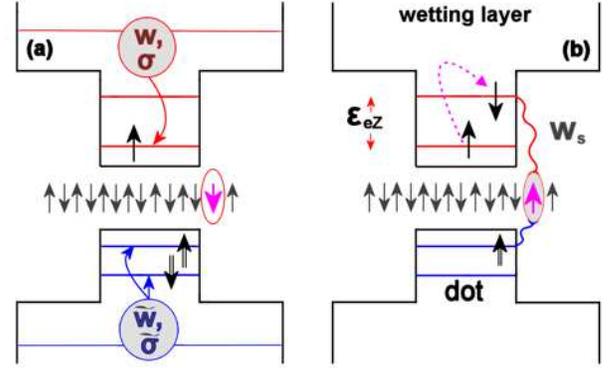}
\caption{\label{schem} (a)Diagram of optical excitation of the dot.
The case of strongly polarized electrons and weakly polarized holes
is shown. (b) Spin-flip-assisted el-h recombination accompanied by
flipping one of the nuclear spins.}
\end{figure}

In this Letter we propose a theory of the nuclear polarization
bistability in optically pumped QDs. We study the dynamics of
nuclear spins in a dot populated by electrons (el) and holes (h)
which arrive into its ground state with the independent rates $w$
(el) and $\tilde{w}$ (h) and polarization degrees $\sigma$ (el) and
$\tilde{\sigma}$ (h) (see Fig.~\ref{schem}(a)). It has been recently
noticed that nuclear polarization bistability may occur in the
regime when light generates 100\% spin-polarized excitons on the
dot. Here, we demonstrate that bistability is a general phenomenon
possible in a wide range of experimental conditions, including the
non-resonant excitation conditions and in the regime when a dot
often appears in a positively charged (trion) state. We also predict
a new phenomenon caused by the bistable behavior of the nuclear spin
orientation: the bistability of the dot average charge.

In optically pumped dots, nuclear spins become orientated due to the
spin flip-flops in which a single electron and one of the nuclei
exchange spins via the hyperfine interaction. The process leading to
spin transfer consists of an el-nucleus spin flip-flop accompanying
the recombination of a polarized electron with a spin $\pm
\tfrac{3}{2}$ heavy hole in a QD carrying a neutral or positively
charged exciton (see Fig.~\ref{schem}(b)). In such a process, the
electron occupies the intermediate inverted-spin state on the dot
virtually, since in a magnetic field a real single-electron
spin-flip is prohibited by energy conservation and the Zeeman
splitting. The rate of the electron spin-flip recombination
involving a single nucleus \cite{erlingsson} is,
\begin{equation}
w_{s}=|u|^2w_{r} / ( \epsilon_{eZ}^2 + \tfrac{1}{4}\gamma^2 )
\label{w_s} .
\end{equation}
Here $u$ is a typical energy of the hyperfine interaction with a
single nucleus, $\gamma$ is broadening of the electron energy level,
and $w_{r}$ is the rate at which the bright exciton recombines on
the dot. The electron Zeeman splitting, modified by the Overhauser
field $B_N$, is $\epsilon_{eZ}=g_{e}\mu_{B}(B-B_{N})$. The form of
the Eq.~(\ref{w_s}) implies a feedback due to the dependence of
$w_s$ on $B_N$, which is key to the nuclear spin bistability.

The kinetic model describing the carrier population in the ground
state of the dot is formulated in terms of the probabilities of its
$16$ allowed configurations based upon the two electron and two hole
spin states corresponding to the lowest el/h orbitals in the QD. We
solve the rate equations for the populations of these states and for
the nuclear orientation, and then find the steady-state magnitude of
the nuclear spin polarization $S$. Here, we denote the probability
that the dot is empty by $n$, and use $n_{\mu}$ ($n^{\mu}$) for the
probabilities of the dot occupation by a single electron (hole),
with the index $\mu=+/-$ representing the spin state of the
particle. We refer to these states as $D, D_{\mu}$ and $D^{\mu}$,
respectively.  The probabilities for the dot to be occupied with two
electrons or two holes (states $D_{+-}$ and $D^{+-}$) are $n_{+-}$
and $n^{+-}$. The probability to find the dot in a dark exciton
state $X_{\mu}^{\mu}$ is $n_{\mu}^{\mu}$, and in a bright exciton
state $X_{\mu}^{-\mu}$ is $n_{\mu}^{-\mu}$. The probability to find
the dot in a negative (positive) trion state labelled as
$X_{+-}^{\mu}$ ($X_{\mu}^{+-}$) is $n_{+-}^{\mu}$ ($n_{\mu}^{+-}$)
and, finally, $n_{+-}^{+-}$ represents the dot in the biexciton
state, $X_{+-}^{+-}$.

Below we list the balance equations for the dot population. The
first two equations describe the probability of the dot occupation
by a single carrier.
\begin{eqnarray}
\dot{n}_{\mu} = && \tfrac{1}{2} (1 + \mu \sigma)   w n + w_r
n_{+-}^{\mu}  -  \left[ \tilde{w} + \tfrac{1}{2} (1 - \mu\sigma)w
\right]n_{\mu};\nonumber
\\
\dot{n}^{\mu}= && \tfrac{1}{2} (1 + \mu \tilde{\sigma}) \tilde{w} n
+ w_r n_{\mu}^{+-} + \tfrac{1}{2} (1+\mu S) Nw_{s} n_{-\mu}^{+-}
\nonumber
\\* && - \left[ w + \tfrac{1}{2} (1 -\mu \tilde{\sigma})\tilde{w} \right]n^{\mu}.
\label{kinset1}
\end{eqnarray}
Both include "gains" due to the arrivals of an electron/hole into
the empty dot (Fig.~\ref{schem}(a)) and the recombination of a
charged bright exciton, and "losses" due to the arrival of an
electron or a hole. The second equation also has a gain due to a
possible spin-flip-assisted recombination from a positive trion
$X_{\mu}^{+-} \rightarrow D^{-\mu}$ in which the spin is transferred
to a nucleus \cite{virtual}, Fig.~\ref{schem}(b). This process is
impossible for a negative trion since in the lowest orbital state
the flip-flop is blocked by the presence of the second electron
\cite{virtual}. The probability for an el-h pair to recombine via
spin-flip depends on the number of nuclei available, which leads to
the term $(1+\mu S) Nw_{s} n_{-\mu}^{+-}$ in Eq.~(\ref{kinset1}),
where $S$ is the degree of nuclear polarization and $N$ is the total
number of nuclei covered by the electron wave function ($N\sim 10^4
\div 10^5$ in a typical InGaAs/GaAs dot).

Equations describing the QD states $D_{+-}$ and $D^{+-}$ are:
\begin{eqnarray}
\dot{n}_{+-}= \tfrac{1}{2}\sum_{\mu}( 1-\mu\sigma) w n_{\mu} -
\tilde{w} n_{+-};\nonumber
\\
\dot{n}^{+-}=  \tfrac{1}{2}\sum_{\mu}( 1 - \mu \tilde{\sigma})
\tilde{w} n^{\mu} - w n^{+-}. \label{kinset2}
\end{eqnarray}

Kinetics of the the neutral bright and dark excitons $X_{\mu}^{\mu}$
and $X_{\mu}^{-\mu}$ are described by
\begin{eqnarray}
\dot{n}_{\mu}^{\mu}=&&  \tfrac{1}{2}(1 + \mu\tilde{\sigma} )
\tilde{w} n_{\mu} + \tfrac{1}{2}(1 + \mu\sigma ) w n^{\mu} \nonumber
\\* &&- \tfrac{1}{2}\left[(1 - \mu S)Nw_{s} + ( 1 - \mu \sigma
) w + ( 1 - \mu \tilde{\sigma} )\tilde{w} \right]
n_{\mu}^{\mu};\nonumber
\\
\dot{n}_{\mu}^{-\mu}= && \tfrac{1}{2}(1 -
\mu\tilde{\sigma})\tilde{w} n_{\mu} + \tfrac{1}{2}(1 + \mu\sigma )w
n^{-\mu} + w_r n_{+-} ^{+-} \nonumber
\\* && -  \left[ w_r + \tfrac{1}{2}(1 - \mu\sigma)w + \tfrac{1}{2}( 1 +
\mu\tilde{\sigma})\tilde{w} \right]n_{\mu}^{-\mu}. \label{kinset3}
\end{eqnarray}
Both neutral bright and dark exciton populations decrease when more
carriers arrive onto the dot. The neutral bright exciton can also be
created and removed due to the el-h pair recombination in the
processes $X^{+-}_{+-}\rightarrow X_{\mu}^{-\mu}$ and
$X_{\mu}^{-\mu}\rightarrow D$, respectively. The dark exciton can
decay due to the spin-flip-assisted recombination ($X_{\mu}^{\mu}
\rightarrow D$) leading to spin transfer to nuclei \cite{virtual}.

Kinetics of the trions $X_{+-}^{\mu}$ and $X_{\mu}^{+-}$ are
described by
\begin{eqnarray}
\dot{n}_{+-}^{\mu} =&&  \tfrac{1}{2}(1 +\mu \tilde{\sigma})
\tilde{w} n_{+-} + \tfrac{1}{2}\sum_{\nu=\pm} (1-\nu
\sigma)wn_{\nu}^{\mu} \nonumber
\\* && -  \left[ w_r + \tfrac{1}{2}(1 - \mu \tilde{\sigma})\tilde{w}
\right] n_{+-}^{\mu} ; \label{kinset4}
\\
\dot{n}_{\mu}^{+-} =&&  \tfrac{1}{2}(1+ \mu \sigma)w n^{+-} +
\tfrac{1}{2}\sum_{\nu=\pm}(1-\nu\sigma)wn_{\mu}^{\nu} \nonumber
\\* && - \left[ w_r + \tfrac{1}{2}(1-\mu S)Nw_{s} +
\tfrac{1}{2}(1-\mu\sigma) w\right] n_{\mu}^{+-}. \nonumber
\end{eqnarray}
Both trion populations change due to the recombinations
$X_{\mu}^{+-} \rightarrow D^{\mu}, X_{+-}^{\mu} \rightarrow D_{\mu}$
and arrival of a single additional charge (the ground states of the
dot permit maximum four carriers). A positive trion can also
recombine in the spin-flip-assisted process $X_{\mu}^{+-}
\rightarrow X^{-\mu}$, forbidden for the negative trions
\cite{virtual}.

Finally, the biexciton state $X_{+-}^{+-}$ cannot contribute to the
nuclear spin pumping as it decays without the spin-flip,
$X_{+-}^{+-}\rightarrow X_{+}^{-},X_{-}^{+}$,
\begin{eqnarray}
\dot{n}_{+-}^{+-} = && \tfrac{1}{2}\sum_{\mu} \left[ (1 -
\mu\tilde{\sigma})\tilde{w} n_{+-}^{\mu} + (1 -\mu\sigma) w
n_{\mu}^{+-} \right]\nonumber
\\* && -  2w_r n_{+-}^{+-}. \label{kinset5}
\end{eqnarray}

The probabilities for the dot with a given nuclear polarization to
be in each of the $16$ configurations are found using the
normalization condition $1=n + n_{+-} + n^{+-} + n_{+-}^{+-}+
\sum_{\mu}n_{\mu}+n^{\mu}+ n_{\mu}^{\mu} +
n_{\mu}^{-\mu}+n_{+-}^{\mu}+n_{\mu}^{+-}$ and the steady state
condition for Eqs.~(\ref{kinset1}-\ref{kinset5}). We formally write
these equations in the form $\hat{M} \vec{n}=\left(
1,0,...,0\right)^T$, where the components of $\vec{n}$ are the
occupation numbers and $\hat{M}$ is a $16\times 16$ matrix with
elements determined by the coefficients in
Eqs.~(\ref{kinset1}-\ref{kinset5}) and the normalization condition.
The formal solutions for components of $\vec{n}$ are given by
$C_{i,1} / detM$ where $C_{i,1}$ is the relevant cofactor of
$\hat{M}$.

A steady-state value for the nuclear polarization S (defined as
$S=f_{\Uparrow}-f_{\Downarrow}$) can be obtained by substituting
formal steady-state solutions of Eqs.~(\ref{kinset1}-\ref{kinset5})
for a given $S$ into the balance equation for the occupation numbers
of spin up ($f_{\Uparrow}$) and down ($f_{\Downarrow}$) nuclei
\cite{spinhalf},
\begin{equation}
\dot{S}=I\equiv \sum_{\mu}\mu\left( 1-\mu S\right)\left(
n_{\mu}^{\mu} + n_{\mu}^{+-} \right)w_{s} - 2Sw_{d}. \label{dotS}
\end{equation}
It summarizes the processes leading to the nuclear spin pumping: $S$
is increased as a result of the spin-flip-assisted recombination of
$X^{+}_{+}$ and $X^{+-}_{+}$ and reduced due to a similar
recombination process involving $X^{-}_{-}$ and $X^{+-}_{-}$. Thus
the balance between the populations of $X^{+}_{+}$ and $X^{+-}_{+}$
on one hand and $X^{-}_{-}$ and $X^{+-}_{-}$ on the other will
eventually define the sign of the net nuclear polarization
\cite{virtual}. However, an additional important contribution to the
depolarization of the nuclei has to be taken into account. It arises
from their mutual dipole-dipole interaction effectively leading to
the nuclear spin diffusion from the dot into the bulk semiconductor
\cite{pagettwo}, described in our model by the rate $w_d$
\cite{dep}.

\begin{figure}
\includegraphics[width=8cm]{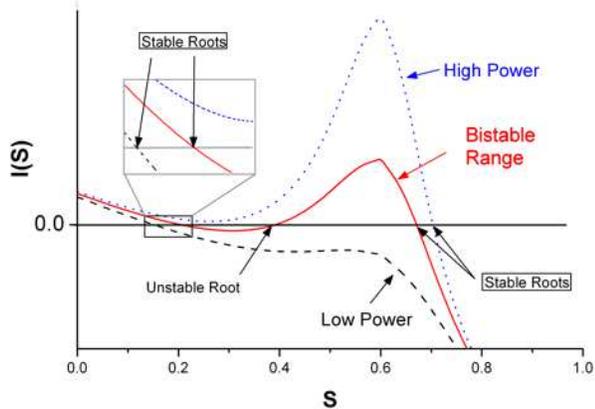}
\caption{\label{I} The function $I(S)$ for the situation where
$w=\tilde{w}, \sigma=0.9, \tilde{\sigma}=-0.2 ,x=0.6$ and $z=8$ for
three different powers: $P=0.0001,0.0003,0.0005$. Stable roots
correspond to the solutions of $I(S)=0$ where $\frac{dI}{dS}<0$.}
\end{figure}

To present the analysis of the above equations, we employ the
following parameters:
\begin{equation}
x=\frac{B}{B_{N}^{max}},\qquad z=2N^2\frac{w_{d}}{w_r},\qquad P=
\frac{\tilde{w}}{zw_r}.\label{P}
\end{equation}
Here $B_N^{max}$ is defined through the Overhauser field $B_{N}$ as
$B_N=B_{N}^{max}S$ and, after introducing $\alpha = \gamma/g_e
\mu_{B}B_{N}^{max}$, Eq.~(\ref{w_s}) can be represented in the form
\cite{values}:
\begin{equation}
w_{s}\equiv \frac{w_{r}}{ N^2\left( x - S\right)^2 +
\tfrac{1}{4}\alpha^2}. \label{w_s2}
\end{equation}
The steady-state values of $S$ determined by the feedback built into
Eqs.~(\ref{w_s}-\ref{w_s2}) are given by the solutions of the
equation $I\left(S\right)=0$, satisfying the condition
$\frac{dI}{dS}<0$ (solutions with $\frac{dI}{dS}>0$ are unstable).
Figure~\ref{I} demonstrates that for a fixed external magnetic field
the number of stable solutions for the nuclear spin polarization
varies: it can be one or two depending on the incident power and
other experimental parameters such as $w_{d},\sigma,\tilde{\sigma}$
and the ratio $w/\tilde{w}$. At small powers only a single low value
of $S$ is possible. At high powers when two stable solutions appear,
including one with a large $S>x$, the dot enters the regime of the
nuclear spin bistability. This result strongly depends on the
depolarization parameter $z$, defined in Eq.~(\ref{P}), so that in
the following discussion we specify the range of $z$ where a
bistability occurs.

\begin{figure}
\includegraphics[width=8cm]{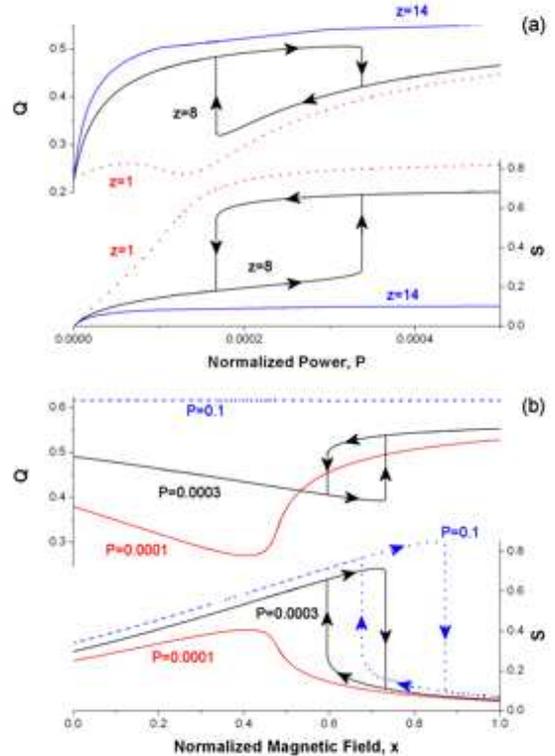}
\caption{\label{hyst} Evolution of nuclear polarization ($S$) and
the average charging state of the dot ($Q$) for
$w=\tilde{w},\sigma=0.9$ and $\tilde{\sigma}=-0.2$: (a) as a
function of power for $x=0.6$ and various values of $z$, with the
arrows indicating a forwards or backwards sweep. Although not shown
in the figure, at high powers $P \approx 1$, both $Q$ and $S$ start
to decrease due to the dot being dominated by the biexciton (for
which the spin-flip process is blocked); (b) as a function of
magnetic field for $z=8$ and various power values.}
\end{figure}

The bottom parts of Figs.~\ref{hyst}(a) and (b) show the calculated
evolution of the nuclear polarization in a dot for realistic
magnitudes of the depolarization parameter $z$ in the regime where
electrons have a high degree of spin memory and arrive with the same
rate as the depolarized holes. Fig.~\ref{hyst}(a) contains a large
hysteresis loop in the power dependence of $S$ for a fixed magnetic
field (here, $x=0.6$), similar to those observed in Refs. 10,11. The
bistable behavior occurs for a wide range of the depolarization
parameter $z$: $5\lesssim z\lesssim 14$. Experimentally, the
evolution of $S$ can be detected in polarization-resolved PL
experiments on individual self-assembled InGaAS/GaAs quantum dots,
by deducing it from the measured exciton Zeeman splitting.

We also find that the bistability in $S$ leads to a novel
phenomenon: a hysteresis in the average dot charge, $Q$ (see top
parts of Fig.~\ref{hyst}(a,b)). This occurs when the electrons
arriving to the dot have a high degree of spin polarization,
permitting their recombination with only one spin orientation of
holes. Thus, an extra hole with the opposite spin is likely to
remain on the dot, leading to, on average, a positive dot charge.
The enhancement of the spin-flip-assisted recombination for large
$S$, removing such holes, will result in reduction of the charge.
Therefore the hysteresis in $S$ will be reflected as a hysteresis in
the average dot charge. A similar bistable behaviour in both $S$ and
$Q$ can also be found if the external magnetic field is varied
\cite{maletinsky} at a fixed optical pump power, as shown in
Fig.~\ref{hyst}(b).

\begin{figure}
\includegraphics[width=8cm]{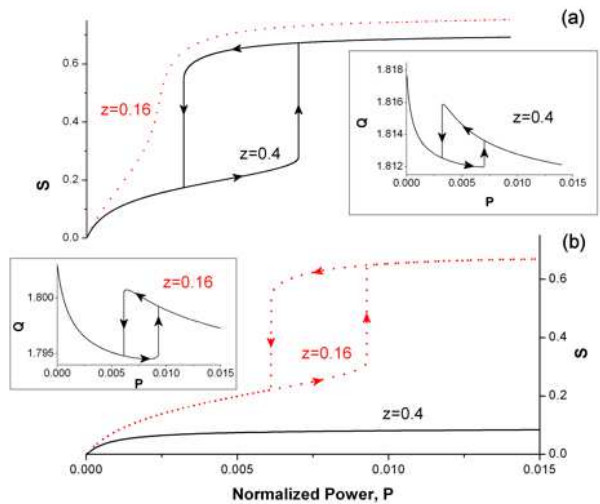}
\caption{\label{diffpol}Evolution of $S$ with $P$ in the regime
where $w=0.1\tilde{w}$ for $z=0.16,0.4$ and different polarizations
of arriving electrons/holes. (a) $\sigma=0.9,\tilde{\sigma}=-0.2$.
The inset shows the evolution of the charging state of the dot for
$z=0.4$ (with a very small hysteresis loop). (b) Same for
$\sigma=0.45,\tilde{\sigma}=-0.1$.}
\end{figure}

Figure~\ref{diffpol} illustrates that the range of parameters for
which the bistability can occur strongly depends on the ratio
between the arrival rates of electrons and holes, $w$ and
$\tilde{w}$, as well as on their polarizations, $\sigma$ and
$\tilde{\sigma}$. In experiment the ratio $w/\tilde{w}$ can be
varied by applying an electric field in a diode containing QDs in
the intrinsic region \cite{braun,lai,tartakovskii,maletinsky}:
because of a light effective mass, electrons can tunnel out before
relaxing to the dot ground state, which in effect reduces their
arrival rate as compared to that of the holes. Fig.~\ref{diffpol}(a)
shows the evolution of $S(P)$ for $w=0.1\tilde{w}$. The dot is
mainly in the state $D^{+-}$ so that its average charge is $Q\approx
+1.8$ and exhibits a weak power-dependence with a negligible
hysteresis loop (see inset), despite a pronounced hysteresis loop in
the nuclear polarization. As seen from the figure, for such low
values of $w/\tilde{w}$ higher powers are required to pump a
significant nuclear polarization, and  the bistability in $S$ is
moved towards smaller values of the depolarization parameter ($0.2
\lesssim z \lesssim 0.5$). Figure~\ref{diffpol}(b) illustrates that
when the polarizations of both electrons and holes is reduced by
$50\%$ the bistability can still be observed, but only for
$0.1\lesssim z \lesssim 0.2$.

To summarize, we have shown that for a wide range of dot parameters
(including the number of nuclei, el-h radiative recombination time
and nuclear spin diffusion rate) the polarization of nuclei in a
non-resonantly optically pumped semiconductor quantum dot can
exhibit a bistable behavior. Thus, we conclude that the nuclear spin
bistability is a general phenomenon for dots pumped with circularly
polarized light. In addition, we find that the nuclear spin
polarization can also strongly influence the charge dynamics in the
dot leading to the bistability of the average dot charge.

We thank A. Imamoglu, O. Tsyplyatyev and A. Yacobi for discussions.
This work has been supported by the Lancaster EPSRC Portfolio
Partnership No. EP/C511743, the Sheffield EPSRC Programme Grant No.
GR/S76076, the EPSRC IRC for Quantum Information Processing,
ESF-EPSRC network Spico EP/D062918, EPSRC Advanced Research
Fellowship EP/C54563X/1.


\begin{thebibliography}{}
\bibitem{overhauser} A. W. Overhauser, Phys. Rev. {\bf 92}, 411 (1953).
\bibitem{pines} D. Pines, J. Bardeen, and C. P. Slichter,
Phys. Rev. {\bf 106}, 489 (1957)
\bibitem{paget} D. Paget {\it et al.}, Phys. Rev. {\bf B 15}, 5780 (1977)
\bibitem{braun} P.-F. Braun {\it et al.}, Phys. Rev. Lett. {\bf 94}, 116601 (2005).
\bibitem{gammon} D. Gammon {\it et al.}, Phys. Rev. Lett. {\bf 86}, 5176 (2001); A. S. Bracker {\it et al.}, {\it ibid.} {\bf 94}, 047402 (2005).
\bibitem{lai} C. W. Lai {\it et al.}, Phys. Rev. Lett. {\bf 96}, 167403 (2006).


\bibitem{yokoi} T. Yokoi {\it et al.}, Phys. Rev. {\bf B 71}, 041307 (2005).

\bibitem{akimov} I. A. Akimov, D. H. Feng and F. Henneberger, Phys. Rev.
Lett. {\bf 97}, 056602 (2006).
\bibitem{eble} B. Eble {\it et al.}, Phys. Rev. {\bf B 74}, 081306 (2006).
\bibitem{brauntwo} P.-F. Braun {\it et al.}, Phys. Rev. {\bf B 74}, 245306 (2006).
\bibitem{tartakovskii} A. I. Tartakovskii {\it et al.}, Phys. Rev. Lett. {\bf 98}, 026806 (2007).
\bibitem{maletinsky} P. Maletinsky {\it et al.}, Phys. Rev. {\bf B 75}, 035409 (2007).

\bibitem{ono} Hints of bistability in nuclear orientation have been seen in the spin-polarized tunneling experiments: K. Ono and S. Tarucha, Phys. Rev. Lett. { \bf 92}, 256803
(2004); S. Tarucha {\it et al.}, Phsy. Stat. Sol. {\bf B 243}, 3673
(2006).
\bibitem{uenoyama} T. Uenoyama and L. J. Sham, Phys. Rev. Lett. {\bf 64},
25 (1989); T. C. Damen {\it et al.}, {\it ibid.} {\bf 67}, 24
(1991).

\bibitem{ebbens} A. Ebbens {\it et al.}, Phys. Rev. {\bf B 72}, 073307 (2005).

\bibitem{virtual} In the opposite polarization of
light, the Overhauser field enhances the Zeeman splitting so that it
cannot cause a bistability. Also, virtual spin-flip processes via
excited orbital states have a much smaller probability, and are thus
neglected. Here we also neglect the effect of the electron-hole
exchange on the electron spin splitting.

\bibitem{erlingsson} S. I. Erlingsson, Y. V. Nazarov, and V. I. Fal'ko,
Phys. Rev. {\bf B 64}, 195306 (2001).
\bibitem{spinhalf} For simplicity, we consider spin $\tfrac{1}{2}$ nuclei. Higher spins will result only in the
re-parameterization of $N$.

\bibitem{pagettwo} D. Paget, Phys. Rev. {\bf B 25}, 4444 (1982).
\bibitem{dep} For a dot with radius $r\approx 5nm$, we approximate
$w_{dep}\approx D_{N}/r^{2}\approx 1-10 s^{-1}$, where $D_{N}\approx
\mu_{n}^{2}/\hbar a$ is the coefficient of polarization diffusion
due to the dipole-dipole interaction between magnetic moments
$\mu_{n}$ of neighbouring nuclei and $a=0.56nm$ is the lattice
constant.

\bibitem{values} For InGaAs quantum dots used in \cite{tartakovskii}, $w_{r}\approx 10^{9}s^{-1}, N\approx 10^4,
B_{N}^{max}\approx 2-3T, \alpha \approx 0.01$.


\end{thebibliography}
\end{document}